\documentclass[a4paper,fleqn]{jpconf}
\usepackage[utf8]{inputenc}
\usepackage[T1]{fontenc}
\usepackage{fouriernc}
\usepackage{amsmath,amsthm,amssymb}
\usepackage{slashed}
\usepackage[all,cmtip]{xy}
\usepackage{multicol}
\newcommand{\ket}[1]{\rvert#1\rangle\,}
\newcommand{\braket}[2]{\,\langle #1\,\rvert\, #2 \rangle \,}
\newcommand{\Braket}[3]{\,\langle #1\,\rvert \, #2 \,\rvert \, #3\, \rangle\,}
\begin{document}
\title{The Bannai-Ito algebra and some applications}

\author{Hendrik De Bie}
\address{Department of Mathematical Analysis, Faculty of Engineering and
Architecture, Ghent University,  Galglaan 2 Galglaan 2, 9000 Ghent, Belgium}
\ead{hendrik.debie@ugent.be}
\author{Vincent X. Genest}
\address{Centre de recherches math\'ematiques, Universit\'e de Montr\'eal, P.O. Box 6128, Centre-ville Station, Montr\'eal (QC) Canada, H3C 3J7}
\ead{vincent.genest@umontreal.ca}
\author{Satoshi Tsujimoto}
\address{Department of applied mathematics and physics, Kyoto University, Kyoto 6068501, Japan}
\ead{tujimoto@i.kyoto-u.ac.jp}
\author{Luc Vinet}
\address{Centre de recherches math\'ematiques, Universit\'e de Montr\'eal, P.O. Box 6128, Centre-ville Station, Montr\'eal (QC) Canada, H3C 3J7}
\ead{luc.vinet@umontreal.ca}
\author{Alexei Zhedanov}
\address{Donetsk Institute for Physics and Technology, Donetsk 340114, Ukraine}
\ead{zhedanov@yahoo.com}

\begin{abstract}
The Bannai-Ito algebra is presented together with some of its applications. Its relations with the Bannai-Ito polynomials, the Racah problem for the $sl_{-1}(2)$ algebra, a superintegrable model with reflections and a Dirac-Dunkl equation on the 2-sphere are surveyed.
\end{abstract}

\section{Introduction}
Exploration through the exact solution of models has a secular tradition in mathematical physics. Empirically, exact solvability is possible in the presence of symmetries, which come in various guises and which are described by a variety of mathematical structures. In many cases, exact solutions are expressed in terms of special functions, whose properties encode the symmetries of the systems in which they arise. This can be represented by the following virtuous circle:
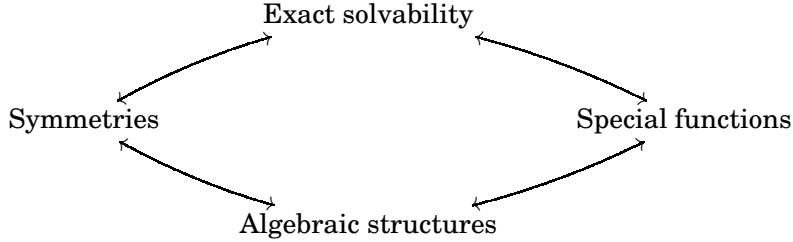
\begin{equation*}
\xymatrix{
  & \text{Exact solvability}  & 
\\
  \text{Symmetries} \ar@/^/@{<->}[ur]&  & \text{Special functions}\ar@/_/@{<->}[ul]
\\
 &  \text{Algebraic structures} \ar@/_/@{<->}[ur]\ar@/^/@{<->}[ul] & 
 }
\end{equation*}
The classical path is the following: start with a model, find its symmetries, determine how these symmetries are mathematically described, work out the representations of that mathematical structure and obtain its relation to special functions to arrive at the solution of the model. However, one can profitably start from any node on this circle. For instance, one can identify and characterize new special functions, determine the algebraic structure they encode, look for models that have this structure as symmetry algebra and proceed to the solution. In this paper, the following path will be taken:
\begin{equation*}
\text{Algebra}\longrightarrow \text{Orthogonal polynomials}\longrightarrow \text{Symmetries}\longrightarrow \text{Exact solutions}
\end{equation*}
The outline of the paper is as follows. In section 2, the Bannai-Ito algebra is introduced and some of its special cases are presented. In section 3, a realization of the Bannai-Ito algebra in terms of discrete shift and reflection operators is exhibited. The Bannai-Ito polynomials and their properties are discussed in section 4. In section 5, the Bannai-Ito algebra is used to derive the recurrence relation satisfied by the Bannai-Ito polynomials. In section 6, the paraboson algebra and the $sl_{-1}(2)$ algebra are introduced. In section 7, the realization of $sl_{-1}(2)$ in terms of Dunkl operators is discussed. In section 8, the Racah problem for $sl_{-1}(2)$ and its relation with the Bannai-Ito algebra is examined. A superintegrable model on the 2-sphere with Bannai-Ito symmetry is studied in section 9. In section 10, a Dunkl-Dirac equation on the 2-sphere with Bannai-Ito symmetry is discussed. A list of open questions is provided in lieu of a conclusion.
\section{The Bannai-Ito algebra}
Throughout the paper, the notation $[A,B]=AB-BA$ and  $\{A,B\}=AB+BA$ will be used. Let $\omega_1$, $\omega_2$ and $\omega_3$ be real parameters. The Bannai-Ito algebra is the associative algebra generated by $K_1$, $K_2$ and $K_3$ together with the three relations
\begin{align}
\label{BI-Algebra}
\{K_1,K_2\}=K_3+\omega_3,\quad \{K_2,K_3\}=K_1+\omega_1,\quad \{K_3,K_1\}=K_2+\omega_2,
\end{align}
or $\{K_i,K_j\}=K_k+\omega_{k}$, with $(ijk)$ a cyclic permutation of $(1,2,3)$. The Casimir operator
\begin{align*}
Q=K_1^2+K_2^2+K_3^2,
\end{align*}
commutes with every generator; this property is easily verified with the commutator identity $[AB,C]=A\{B,C\}-\{A,C\}B$. Let us point out two special cases of \eqref{BI-Algebra} that have been considered previously in the literature.
\begin{enumerate}
\item $\omega_1=\omega_2=\omega_3=0$
\end{enumerate}
The special case with defining relations
\begin{align*}
\{K_1,K_2\}=K_3,\quad \{K_2,K_3\}=K_1,\quad \{K_3,K_1\}=K_2,
\end{align*}
is sometimes referred to as the \emph{anticommutator spin algebra} \cite{Arik-2003, Gorodnii-1984}; representations of this algebra were examined in \cite{Arik-2003, Gorodnii-1984, Brown-2013, Silvestrov-1992}.
\begin{enumerate}\setcounter{enumi}{1}
\item $\omega_1=\omega_2=0\neq \omega_3$
\end{enumerate}
In recent work on the construction of novel finite oscillator models \cite{Jafarov-2011-05, VDJ-2011}, E. Jafarov, N. Stoilova and J. Van der Jeugt introduced the following extension of $\mathfrak{u}(2)$
by an involution $R$ ($R^2=1$):
\begin{gather*}
[I_3,R]=0,\quad \{I_1,R\}=0,\quad \{I_2,R\}=0,
\\
[I_3,I_1]=i I_2,\quad [I_2,I_3]=i I_1,\quad [I_1,I_2]=i (I_3+\omega_3 R).
\end{gather*}
It is easy to check that with 
\begin{align*}
K_1=i I_1 R,\quad K_2=I_2,\quad K_3=I_3 R,
\end{align*}
the above relations are converted into
\begin{align*}
\{K_1,K_3\}=K_2,\quad \{K_2,K_3\}= K_1,\quad \{K_1,K_2\}=K_3+\omega_3.
\end{align*}
\section{A realization of the Bannai-Ito algebra with shift and reflections operators}
Let $T^{+}$ and $R$ be defined as follows:
\begin{align*}
T^{+}f(x)=f(x+1),\quad Rf(x)=f(-x).
\end{align*}
Consider the operator 
\begin{align}
\label{K1-Hat}
\widehat{K}_1=F(x) (1-R)+G(x)(T^{+}R-1)+h,\qquad h=\rho_1+\rho_2-r_1-r_2+1/2,
\end{align}
with $F(x)$ and $G(x)$ given by
\begin{gather*}
F(x)=\frac{(x-\rho_1)(x-\rho_2)}{x},\quad G(x)=\frac{(x-r_1+1/2)(x-r_2+1/2)}{x+1/2},
\end{gather*}
where $\rho_1, \rho_2,r_1, r_2$ are four real parameters. It can be shown that $\widehat{K}_1$ is the most general operator of first order in $T^{+}$ and $R$ that stabilizes the space of polynomials of a given degree \cite{Tsujimoto-2012-03}. That is, for any polynomial $Q_{n}(x)$ of degree $n$, $[\widehat{K}_1 Q_{n}(x)]$ is also a polynomial of degree $n$. Introduce
\begin{align}
\label{K2-Hat}
\widehat{K}_2=2x+1/2,
\end{align}
which is essentially the ``multiplication by $x$'' operator and
\begin{align}
\label{K3-Hat}
\widehat{K}_3\equiv \{\widehat{K}_1,\widehat{K}_2\}-4(\rho_1\rho_2-r_1r_2).
\end{align}
It is directly verified that $\widehat{K}_1$, $\widehat{K}_2$ and $\widehat{K}_3$ satisfy the commutation relations
\begin{align}
\label{BI-Hat}
\{\widehat{K}_1,\widehat{K}_2\}=\widehat{K}_3+\widehat{\omega}_3,\quad \{\widehat{K}_2,\widehat{K}_3\}=\widehat{K}_1+\widehat{\omega}_1,\quad \{\widehat{K}_3,\widehat{K}_1\}=\widehat{K}_2+\widehat{\omega}_2,
\end{align}
where the structure constants $\widehat{\omega}_1$, $\widehat{\omega}_2$ and $\widehat{\omega}_3$ read
\begin{align}
\label{Omega-Hat}
\widehat{\omega}_1=4(\rho_1\rho_2+r_1r_2),\quad \widehat{\omega}_2=2(\rho_1^2+\rho_2^2-r_1^2-r_2^2),\quad \widehat{\omega}_3=4(\rho_1\rho_2-r_1r_2).
\end{align}
The operators $\widehat{K}_1$, $\widehat{K}_2$ and $\widehat{K}_3$ thus realize the Bannai-Ito algebra. In this realization, the Casimir operator acts as a multiple of the identity; one has indeed
\begin{align*}
\widehat{Q}=\widehat{K}_1^2+\widehat{K}_2^2+\widehat{K}_3^2=2(\rho_1^2+\rho_2^2+r_1^2+r_2^2)-1/4.
\end{align*}
\section{The Bannai-Ito polynomials}
Since the operator \eqref{K1-Hat} preserves the space of polynomials of a given degree, it is natural to look for its eigenpolynomials, denoted by $B_{n}(x)$, and their corresponding eigenvalues $\lambda_{n}$. We use the following notation for the generalized hypergeometric series \cite{Andrews_Askey_Roy_1999}
\begin{align*}
{}_rF_{s}\left(\genfrac{}{}{0pt}{}{a_1,\ldots, a_{r}}{b_{1},\ldots,b_{s}}\,\Big \rvert \, z\right)=\sum_{k=0}^{\infty}\frac{(a_1)_{k}\cdots (a_{r})_{k}}{(b_1)_{k}\cdots (b_{s})_{k}}\,\frac{z^{k}}{k!},
\end{align*}
where $(c)_{k}=c(c+1)\cdots (c+k-1)$, $(c)_0\equiv 1$ stands for the Pochhammer symbol; note that the above series terminates if one of the $a_{i}$ is a negative integer. Solving the eigenvalue equation
\begin{align}
\label{BI-Eigen}
\widehat{K}_1 B_{n}(x)=\lambda_{n}B_{n}(x),\qquad n=0,1,2,\ldots
\end{align}
it is found that the eigenvalues $\lambda_{n}$ are given by \cite{Tsujimoto-2012-03}
\begin{align}
\label{BI-Eigenvalues}
\lambda_{n}=(-1)^{n}(n+h),
\end{align}
and that the polynomials have the expression
\begin{align}
\frac{B_{n}(x)}{c_{n}}=
\begin{cases}
{}_{4}F_{3}\left(\genfrac{}{}{0pt}{}{-\frac{n}{2},\,\frac{n+1}{2}+h,\, x-r_1+1/2,\,-x-r_1+1/2}{1-r_1-r_2,\, \rho_1-r_1+\frac{1}{2},\, \rho_2-r_1+\frac{1}{2}}\,\Big \rvert \, 1\right) & 
\\[.5cm]
\quad +\frac{(\frac{n}{2})(x-r_1+\frac{1}{2})}{(\rho_1-r_1+\frac{1}{2})(\rho_2-r_1+\frac{1}{2})}\;{}_{4}F_{3}\left(\genfrac{}{}{0pt}{}{1-\frac{n}{2},\,\frac{n+1}{2}+h,\,x-r_1+3/2,\,-x-r_1+1/2}{1-r_1-r_2,\,\rho_1-r_1+\frac{3}{2},\, \rho_2-r_1+\frac{3}{2}}\,\Big\rvert \, 1\right) & \text{$n$ even},
\\[.7cm]
{}_4F_{3}\left(\genfrac{}{}{0pt}{}{-\frac{n-1}{2},\, \frac{n}{2}+h,\, x-r_1+\frac{1}{2},\, -x-r_1+\frac{1}{2}}{1-r_1-r_2,\, \rho_1-r_1+\frac{1}{2},\, \rho_2-r_1+\frac{1}{2}}\,\Big\rvert\,1\right) &
\\[.5cm]
\quad - \frac{(\frac{n}{2}+h)(x-r_1+\frac{1}{2})}{(\rho_1-r_1+\frac{1}{2})(\rho_2-r_1+\frac{1}{2})} \;
{}_4F_{3}\left(\genfrac{}{}{0pt}{}{-\frac{n-1}{2},\, \frac{n+2}{2}+h,\, x-r_1+\frac{3}{2},\, -x-r_1+\frac{1}{2}}{1-r_1-r_2,\, \rho_1-r_1+\frac{3}{2},\, \rho_2-r_1+\frac{3}{2}}\,\Big \rvert \, 1 \right)
& \text{$n$ odd},
\end{cases}
\label{BI-OPs}
\end{align}
where the coefficient
\begin{align*}
c_{2n+p}&=(-1)^{p}\frac{(1-r_1-r_2)_{n}(\rho_1-r_1+1/2,\rho_2-r_1+1/2)_{n+p}}{(n+h+1/2)_{n+p}},\qquad p\in \{0,1\},
\end{align*}
ensures that the polynomials $B_{n}(x)$ are monic, i.e. $B_{n}(x)=x^{n}+\mathcal{O}(x^{n-1})$. The polynomials \eqref{BI-OPs} were first written down by Bannai and Ito in their classification of the orthogonal polynomials satisfying the \emph{Leonard duality} property \cite{Leonard-1982-07, Bannai-1984}, i.e. polynomials $p_{n}(x)$ satisfying both
\begin{itemize}
\item A 3-term recurrence relation with respect to the degree $n$,
\item A 3-term difference equation with respect to a variable index $s$.
\end{itemize}
The identification of the defining eigenvalue equation \eqref{BI-Eigen} of the Bannai-Ito polynomials in \cite{Tsujimoto-2012-03} has allowed to develop their theory. That they obey a three-term difference equation stems from the fact that there are grids such as
\begin{align*}
x_{s}=(-1)^{s}(s/2+a+1/4)-1/4,
\end{align*}
for which operators of the form
\begin{align*}
H=A(x) R+B(x) T^{+}R+C(x),
\end{align*}
are tridiagonal in the basis $f(x_s)$
\begin{align*}
Hf(x_{s})=
\begin{cases}
B(x_{s}) f(x_{s+1})+ A(x_{s}) f(x_{s-1})+C(x_{s}) f(x_{s}) & \text{$s$ even},
\\
A(x_{s}) f(x_{s+1})+B(x_{s})f(x_{s-1})+C(x_{s})f(x_{s}) & \text{$s$ odd}.
\end{cases}
\end{align*}
It was observed by Bannai and Ito that the polynomials \eqref{BI-OPs} correspond to a $q\rightarrow -1$ limit of the $q$-Racah polynomials (see \cite{Koekoek-2010} for the definition of $q$-Racah polynomials). In this connection, it is worth mentioning that the Bannai-Ito algebra \eqref{BI-Hat} generated by the defining operator $\widehat{K}_1$ and the recurrence operator $\widehat{K}_2$ of the 
Bannai-Ito polynomials can be obtained as a $q\rightarrow -1$ limit of the Zhedanov algebra \cite{Zhedanov-1991-11}, which encodes the bispectral property of the $q$-Racah polynomials. The Bannai-Ito polynomials $B_{n}(x)$ have companions
\begin{align*}
I_{n}(x)=\frac{B_{n+1}(x)-\frac{B_{n+1}(\rho_1)}{B_{n}(\rho_1)}B_{n}(x)}{x-\rho_1},
\end{align*}
called the \emph{complementary} Bannai-Ito polynomials \cite{Genest-2013-02-1}. It has now been understood that the polynomials $B_{n}(x)$ and $I_{n}(x)$ are the ancestors of a rich ensemble of polynomials referred to as ``$-1$ orthogonal polynomials'' \cite{Tsujimoto-2012-03,Genest-2013-02-1, Genest-2013-09-02, Vinet-2011-01,Vinet-2012-05, Tsujimoto-2013-03-01, Vinet-2011}. All polynomials of this scheme are eigenfunctions of first or second order operators of Dunkl type, i.e. which involve reflections.
\section{The recurrence relation of the BI polynomials from the BI algebra}
Let us now show how the Bannai-Ito algebra can be employed to derive the recurrence relation satisfied by the Bannai-Ito polynomials. In order to obtain this relation, one needs to find the action of the operator $\widehat{K}_2$ on the BI polynomials $B_{n}(x)$. Introduce the operators
\begin{align}
\label{KPM}
\begin{aligned}
\widehat{K}_{+}=(\widehat{K}_2+\widehat{K}_3)(\widehat{K}_1-1/2)-\frac{\widehat{\omega}_2+\widehat{\omega}_3}{2},\quad 
\widehat{K}_{-}=(\widehat{K}_2-\widehat{K}_3)(\widehat{K}_1+1/2)+\frac{\widehat{\omega}_2-\widehat{\omega}_3}{2},
\end{aligned}
\end{align}
where $\widehat{K}_i$ and $\widehat{\omega}_i$ are given by \eqref{K1-Hat}, \eqref{K2-Hat}, \eqref{K3-Hat} and \eqref{Omega-Hat}. It is readily checked using \eqref{BI-Hat} that
\begin{align*}
\{\widehat{K}_1,\widehat{K}_{\pm}\}=\pm K_{\pm}.
\end{align*}
One can directly verify that $\widehat{K}_{\pm}$ maps polynomials to polynomials. In view of the above, one has
\begin{align*}
\widehat{K}_1 \widehat{K}_{+} B_{n}(x)=(-\widehat{K}_{+}\widehat{K}_1+\widehat{K}_{+})B_{n}(x)=(1-\lambda_{n})\widehat{K}_{+}B_{n}(x),
\end{align*}
where $\lambda_{n}$ is given by \eqref{BI-Eigenvalues}. It is also seen from \eqref{BI-Eigenvalues} that 
\begin{align*}
1-\lambda_{n}=
\begin{cases}
\lambda_{n-1} & \text{$n$ even},
\\
\lambda_{n+1} & \text{$n$ odd}.
\end{cases}
\end{align*}
It follows that
\begin{align*}
\widehat{K}_{+}B_{n}(x)=
\begin{cases}
\alpha_{n}^{(0)}B_{n-1}(x) & \text{$n$ even},
\\
\alpha_{n}^{(1)} B_{n+1}(x) & \text{$n$ odd}.
\end{cases}
\end{align*}
Similarly, one finds
\begin{align*}
\widehat{K}_{-}B_{n}(x)=
\begin{cases}
\beta_{n}^{(0)} B_{n+1}(x) & \text{$n$ even},
\\
\beta_{n}^{(1)} B_{n-1}(x) & \text{$n$ odd}.
\end{cases}
\end{align*}
The coefficients
\begin{gather*}
\alpha_{n}^{(0)}=\frac{2n(\frac{n}{2}+\rho_1+\rho_2)(r_1+r_2-\frac{n}{2})(\frac{n-1}{2}+h)}{n+h-\frac{1}{2}},\quad \alpha_{n}^{(1)}=-4(n+h+1/2),
\\
\beta_{n}^{(0)}=4(n+h+1/2),\quad \beta_{n}^{(1)}=\frac{4(\rho_1-r_1+\frac{n}{2})(\rho_2-r_1+\frac{n}{2})(\rho_1-r_2+\frac{n}{2})(\rho_2-r_2+\frac{n}{2})}{n+h-1/2},
\end{gather*}
can be obtained from the comparison of the highest order term. Introduce the operator
\begin{align}
\label{V-1}
V=\widehat{K}_{+}(\widehat{K}_1+1/2)+\widehat{K}_{-}(\widehat{K}_1-1/2).
\end{align}
From the definition \eqref{KPM} of $\widehat{K}_{\pm}$, it follows that 
\begin{align}
\label{V-2}
V=2 \widehat{K}_2(\widehat{K}_1^2-1/4)-\widehat{\omega}_3 \widehat{K}_1-\widehat{\omega}_2/2.
\end{align}
From \eqref{BI-Eigen},  \eqref{V-1} and the actions of the operators $\widehat{K}_{\pm}$, we find that $V$ is two-diagonal
\begin{align}
\label{First}
V B_{n}(x)=
\begin{cases}
(\lambda_n+1/2) \alpha_{n}^{(0)} B_{n-1}(x)+(\lambda_{n}-1/2)\beta_{n}^{(0)} B_{n+1}(x) & \text{$n$ even},
\\
(\lambda_n-1/2)\beta_{n}^{(1)} B_{n-1}(x)+(\lambda_{n}+1/2)\alpha_{n}^{(1)} B_{n+1}(x) & \text{$n$ odd}.
\end{cases}
\end{align}
From \eqref{V-2} and recalling the definition \eqref{K2-Hat} of $\widehat{K}_2$, we have also
\begin{align}
\label{Second}
V B_{n}(x)=\left[(\lambda_n^2-1/4)(4x+1)-\widehat{\omega}_3 \lambda_n-\widehat{\omega}_2/2\right]B_{n}(x).
\end{align}
Upon combining \eqref{First} and \eqref{Second}, one finds that the Bannai-Ito polynomials satisfy the three-term recurrence relation
\begin{align*}
x\,B_{n}(x)=B_{n+1}(x)+(\rho_1-A_{n}-C_{n}) B_{n}(x)+A_{n-1}C_{n} B_{n-1}(x),
\end{align*}
where
\begin{align}
\label{BI-RECU}
\begin{aligned}
A_{n}&=
\begin{cases}
\frac{(n+1+2\rho_1-2r_1)(n+1+2\rho_1-2r_2)}{4(n+\rho_1+\rho_2-r_1-r_2+1)} & \text{$n$ even},
\\
\frac{(n+1+2\rho_1+2\rho_2-2r_1-2r_2)(n+1+2\rho_1+2\rho_2)}{4(n+\rho_1+\rho_2-r_1-r_2+1)} & \text{$n$ odd},
\end{cases}
\\
C_{n}&=
\begin{cases}
-\frac{n(n-2r_1-2r_2)}{4(n+\rho_1+\rho_2-r_1-r_2)} & \text{$n$ even},
\\
-\frac{(n+2\rho_2-2r_2)(n+2\rho_2-2r_1)}{4(n+\rho_1+\rho_2-r_1-r_2)} & \text{$n$ odd}.
\end{cases}
\end{aligned}
\end{align}
The positivity of the coefficient $A_{n-1}C_{n}$ restricts the polynomials $B_{n}(x)$ to being orthogonal on a finite set of points \cite{Chihara-2011}.
\section{The paraboson algebra and $sl_{-1}(2)$}
The next realization of the Bannai-Ito algebra will involve $sl_{-1}(2)$; this algebra, introduced in \cite{Tsujimoto-2011-10}, is closely related to the parabosonic oscillator.
\subsection{The paraboson algebra}
Let $a$ and $a^{\dagger}$ be the generators of the paraboson algebra. These generators satisfy \cite{Green-1953}
\begin{align*}
[\{a,a^{\dagger}\}, a]=-2a,\quad [\{a,a^{\dagger}\},a^{\dagger}]=2a^{\dagger}.
\end{align*}
Setting $H=\frac{1}{2}\{a,a^{\dagger}\}$, the above relations amount to
\begin{align*}
[H, a]=-a,\quad [H, a^{\dagger}]=a^{\dagger},
\end{align*}
which correspond to the quantum mechanical equations of an oscillator.
\subsection{Relation with $\mathfrak{osp}(1|2)$}
The paraboson algebra is related to the Lie superalgebra $\mathfrak{osp}(1|2)$ \cite{Ganchev-1980}. Indeed, upon setting
\begin{align*}
F_{-}=a,\quad F_{+}=a^{\dagger},\quad E_0=H=\frac{1}{2}\{F_{+}, F_{-}\},\quad E_{+}=\frac{1}{2} F_{+}^2,\quad E_{-}=\frac{1}{2} F_{-}^2,
\end{align*}
and interpreting $F_{\pm}$ as odd generators, it is directly verified that the generators $F_{\pm}$, $E_{\pm}$ and $E_0$ satisfy the defining relations of $\mathfrak{osp}(1|2)$ \cite{Kac-1977}:
\begin{gather*}
[E_0, F_{\pm}]=\pm F_{\pm},\quad \{F_{+}, F_{-}\}=2 E_0,\quad [E_0, E_{\pm}]=\pm 2E_{\pm},\quad [E_{-}, E_{+}]=E_0, 
\\
[F_{\pm}, E_{\pm}]=0,\quad [F_{\pm}, E_{\mp}]=\mp F_{\mp}.
\end{gather*}
The $\mathfrak{osp}(1|2)$ Casimir operator reads
\begin{align*}
C_{\mathfrak{osp}(1|2)}=(E_0-1/2)^2-4E_{+}E_{-}-F_{+}F_{-}.
\end{align*}
\subsection{$sl_{q}(2)$}
Consider now the quantum algebra $sl_{q}(2)$. It can be presented in terms of the generators $A_{0}$ and $A_{\pm}$ satisfying the commutation relations \cite{Vilenkin-1991}
\begin{align*}
[A_0,A_{\pm}]=\pm A_{\pm},\quad [A_{-}, A_{+}]=2 \frac{q^{A_0}-q^{-A_0}}{q-q^{-1}}.
\end{align*}
Upon setting
\begin{align*}
B_{+}=A_{+} q^{(A_0-1)/2},\quad B_{-}=q^{(A_0-1)/2}A_{-},\quad B_0=A_0,
\end{align*}
these relations become
\begin{align*}
[B_0, B_{\pm}]=\pm B_{\pm},\quad B_{-}B_{+}-q B_{+}B_{-}=2\frac{q^{2 B_0}-1}{q^2-1}.
\end{align*}
The $sl_{q}(2)$ Casimir operator is of the form
\begin{align*}
C_{sl_{q}(2)}=B_{+}B_{-}q^{-B_0}-\frac{2}{(q^2-1)(q-1)}(q^{B_0-1}+q^{-B_0}).
\end{align*}
Let $j$ be a non-negative integer. The algebra $sl_{q}(2)$ admits a discrete series representation on the basis $\ket{j,n}$ with the actions
\begin{align*}
q^{B_0}\ket{j,n}=q^{j+n}\ket{j,n},\qquad n=0,1,2,\ldots.
\end{align*} 
The algebra has a non-trivial coproduct $\Delta: sl_{q}(2)\rightarrow sl_{q}(2)\otimes sl_{q}(2)$ which reads
\begin{align*}
\Delta(B_0)=B_0\otimes 1+1\otimes B_0,\quad \Delta(B_{\pm})=B_{\pm}\otimes q^{B_0}+1\otimes B_{\pm}.
\end{align*}
\subsection{The $sl_{-1}(2)$ algebra as a $q\rightarrow -1$ limit of $sl_{q}(2)$}
The $sl_{-1}(2)$ algebra can be obtained as a $q\rightarrow -1$ limit of $sl_{q}(2)$. Let us first introduce the operator $R$ defined as
\begin{align*}
R=\lim_{q\rightarrow -1} q^{B_{0}}.
\end{align*}
It is easily seen that 
\begin{align*}
R\ket{j,n}=(-1)^{j+n}\ket{j,n}=\epsilon (-1)^{n} \ket{j,n},
\end{align*}
where $\epsilon=\pm 1$ depending on the parity of $j$, thus $R^2=1$. When $q\rightarrow -1$, one finds that
\begin{gather*}
\begin{matrix}
q^{B_0}B_{+}=q B_{+}q^{B_0}
\\
B_{-}q^{B_0}=q q^{B_0} B_{-}
\end{matrix}
\longrightarrow
\{R, B_{\pm}\}=0,
\\
B_{-}B_{+}-q B_{+}B_{-}=2\frac{q^{2B_0}-1}{q^2-1}\longrightarrow \{B_{+}, B_{-}\}=2B_{0},
\\
C_{sl_{q}(2)}\longrightarrow B_{+}B_{-}R-B_{0}R+R/2,
\\
\Delta(B_{\pm})=B_{\pm}\otimes q^{B_0}+1\otimes B_{\pm}\longrightarrow \Delta(B_{\pm})=B_{\pm}\otimes R+1\otimes B_{\pm}.
\end{gather*}
In summary, $sl_{-1}(2)$ is the algebra generated by $J_{0}$, $J_{\pm}$ and $R$ with the relations \cite{Tsujimoto-2011-10}
\begin{align}
\label{SL}
[J_0, J_{\pm}]=\pm J_{\pm},\quad [J_0, R]=0,\quad \{J_{\pm}, R\}=0,\quad \{J_{+}, J_{-}\}=2 J_0,\quad R^2=1.
\end{align}
The Casimir operator has the expression
\begin{align}
\label{SL-Casimir}
Q=J_{+}J_{-}R-J_0 R+R/2,
\end{align}
and the coproduct is of the form \cite{Daska-2000-02} 
\begin{align}
\label{SL-Coproduct}
\Delta(J_0)=J_0\otimes 1+1\otimes J_0,\quad \Delta(J_{\pm})=J_{\pm}\otimes R+1\otimes J_{\pm},\quad \Delta(R)=R\otimes R.
\end{align}
The $sl_{-1}(2)$ algebra \eqref{SL} has irreducible and unitary discrete series representations with basis $\ket{\epsilon, \mu;n}$, where $n$ is a non-negative integer,  $\epsilon=\pm 1$ and $\mu$ is a real number such that $\mu>-1/2$. These representations are defined by the following actions:
\begin{gather*}
J_{0}\ket{\epsilon,\mu;n}=(n+\mu+\frac{1}{2})\ket{\epsilon, \mu;n},\quad R\ket{\epsilon,\mu;n}=\epsilon (-1)^{n}\ket{\epsilon,\mu;n},
\\
J_{+}\ket{\epsilon,\mu;n}=\rho_{n+1}\ket{\epsilon,\mu;n+1},\quad J_{-}\ket{\epsilon,\mu;n}=\rho_{n}\ket{\epsilon,\mu;n-1},
\end{gather*}
where $\rho_{n}=\sqrt{n+\mu (1-(-1)^{n})}$. In these representations, the Casimir operator takes the value
\begin{align*}
Q\ket{\epsilon,\mu;n}=-\epsilon \mu \ket{\epsilon,\mu;n}.
\end{align*}
These modules will be denoted by $V^{(\epsilon,\mu)}$. Let us offer the following remarks.
\begin{itemize}
\item  The $sl_{-1}(2)$ algebra corresponds to the parabose algebra supplemented by $R$. 
\item The $sl_{-1}(2)$ algebra consists of the Cartan generator $J_0$ and the two odd elements of $\mathfrak{osp}(1|2)$ supplemented by the involution $R$.
\item One has $C_{\mathfrak{osp}(1|2)}=Q^2$, where $Q$ is given by \eqref{SL-Casimir}. Thus the introduction of $R$ allows to take the square-root of $C_{\mathfrak{osp}(1|2)}$.
\item In $sl_{-1}(2)$, one has $[J_{-}, J_{+}]=1-2 QR$. On the module $V^{(\epsilon,\mu)}$, this leads to
\begin{align*}
[J_{-}, J_{+}]=1+2\epsilon \mu R.
\end{align*}
\end{itemize}
\section{Dunkl operators}
The irreducible modules $V^{(\epsilon,\mu)}$ of $sl_{-1}(2)$ can be realized by Dunkl operators on the real line. Let $R_{x}$ be the reflection operator 
\begin{align*}
R_{x} f(x)=f(-x).
\end{align*}
The $\mathbb{Z}_2$-Dunkl operator on $\mathbb{R}$ is defined by \cite{Dunkl-1989-01}
\begin{align*}
D_{x}=\frac{\partial}{\partial x}+\frac{\nu}{x}(1-R_{x}),
\end{align*}
where $\nu$ is a real number such that $\nu>-1/2$. Upon introducing the operators
\begin{align*}
\widehat{J}_{\pm}=\frac{1}{\sqrt{2}}(x\mp D_{x}),
\end{align*}
and defining $\widehat{J}_0=\frac{1}{2}\{\widehat{J}_{-}, \widehat{J}_{+}\}$, it is readily verified that a realization of the $sl_{-1}(2)$-module $V^{(\epsilon,\mu)}$ with $\epsilon=1$ and $\mu=\nu$ is obtained. In particular, one has
\begin{align*}
[\widehat{J}_{-}, \widehat{J}_{+}]=1+2\nu R_{x}.
\end{align*}
It can be seen that $\widehat{J}_{\pm}^{\dagger}=\widehat{J}_{\mp}$ with respect to the measure $|x|^{2\nu}\,\mathrm{d}x$ on the real line \cite{Genest-2013-04}.
\section{The Racah problem for $sl_{-1}(2)$ and the Bannai-Ito algebra}
The Racah problem for $sl_{-1}(2)$ presents itself when the direct product of three irreducible representations is examined. We consider the three-fold tensor product
\begin{align*}
V=V^{(\epsilon_1,\mu_1)}\otimes V^{(\epsilon_2,\mu_2)}\otimes V^{(\epsilon_3,\mu_3)}.
\end{align*}
It follows from the coproduct formula \eqref{SL-Coproduct} that the generators of $sl_{-1}(2)$ on $V$ are of the form
\begin{gather*}
J^{(4)}=J_0^{(1)}+J_0^{(2)}+J_0^{(3)},\quad J_{\pm}^{(4)}=J_{\pm}^{(1)}R^{(2)}R^{(3)}+J_{\pm}^{(2)}R^{(3)}+J_{\pm}^{(3)},\quad R^{(4)}=R^{(1)}R^{(2)}R^{(3)},
\end{gather*}
where the superscripts indicate on which module the generators act. In forming the module $V$, two sequences are possible: one can first combine $(1)$ and $(2)$ to bring $(3)$ after or one can combine $(2)$ and $(3)$ before adding $(1)$. This is represented by
\begin{align}
\label{Coupling-Scheme}
\left(V^{(\epsilon_1,\mu_1)}\otimes V^{(\epsilon_2,\mu_2)}\right)\otimes V^{(\epsilon_3,\mu_3)}\quad \text{or}\quad V^{(\epsilon_1,\mu_1)}\otimes \left(V^{(\epsilon_2,\mu_2)}\otimes V^{(\epsilon_3,\mu_3)}\right).
\end{align}
These two addition schemes are equivalent and the two corresponding bases are unitarily related. In the following, three types of Casimir operators will be  distinguished. 
\begin{itemize}
\item The initial Casimir operators
\begin{align*}
Q_{i}=J_{+}^{(i)}J_{-}^{(i)}R^{(i)}-(J_0^{(i)}-1/2)R^{(i)}=-\epsilon_i\mu_i,\qquad i=1,2,3.
\end{align*}
\item The intermediate Casimir operators
\begin{align*}
Q_{ij}&=(J_{+}^{(i)}R^{(j)}+J_{+}^{(j)})(J_{-}^{(i)}R^{(j)}+J_{-}^{(j)})R^{(i)}R^{(j)}-(J_{0}^{(i)}+J_0^{(j)}-1/2)R^{(i)}R^{(j)}
\\
&=(J_{-}^{(i)}J_{+}^{(j)}-J_{+}^{(i)}J_{-}^{(j)})R^{(i)}-R^{(i)}R^{(j)}/2+Q_{i}R^{(j)}+Q_{j}R^{(i)},
\end{align*}
where $(ij)=(12), (23)$.
\item The total Casimir operator
\begin{align*}
Q_{4}=[J_{+}^{(4)}J_{-}^{(4)}-(J_0^{(4)}-1/2)]R^{(4)}.
\end{align*}
\end{itemize}
Let $\ket{q_{12}, q_{4}; m}$ and $\ket{q_{23}, q_{4};m}$ be the orthonormal bases associated to the two coupling schemes presented  in \eqref{Coupling-Scheme}. These two bases are defined by the relations
\begin{align*}
Q_{12}\ket{q_{12}, q_{4};m}=q_{12}\ket{q_{12}, q_{4};m},\quad Q_{23}\ket{q_{23},q_{4};m}=q_{23}\ket{q_{23},q_{4};m},
\end{align*}
and
\begin{align*}
Q_{4} \ket{-, q_{4};m}=q_{4} \ket{-, q_{4};m},\quad J_0^{(4)}\ket{-, q_{4};m}=(m+\mu_1+\mu_2+\mu_3+3/2)\ket{-,q_{4};m}.
\end{align*}
The Racah problem consists in finding the overlap coefficients
\begin{align*}
\braket{q_{23}, q_{4}}{q_{12},q_{4}},
\end{align*}
between the eigenbases of $Q_{12}$ and $Q_{23}$ with a fixed value $q_{4}$ of the total Casimir operator $Q_{4}$; as these coefficients do not depend on $m$, we drop this label. For simplicity, let us now take
\begin{align*}
\epsilon_1=\epsilon_2=\epsilon_3=1.
\end{align*}
Upon defining
\begin{align*}
K_1=-Q_{23},\qquad K_3=-Q_{12},
\end{align*}
one finds that the intermediate Casimir operators of $sl_{-1}(2)$ realize the Bannai-Ito algebra \cite{Genest-2012}
\begin{align}
\label{BI-Racah}
\{K_1,K_3\}=K_2+\Omega_2,\quad \{K_1,K_2\}=K_3+\Omega_3,\quad \{K_2,K_3\}=K_1+\Omega_1,
\end{align}
with structure constants
\begin{align}
\label{Structure-Constants}
\Omega_1=2(\mu_1\mu+\mu_2\mu_3),\quad \Omega_2=2(\mu_1\mu_3+\mu_2 \mu),\quad \Omega_3=2(\mu_1\mu_2+\mu_3\mu),
\end{align}
where $\mu=\epsilon_4 \mu_4=-q_{4}$. The first relation in \eqref{BI-Racah} can be taken to define $K_2$ which reads
\begin{align*}
K_2=(J_{+}^{(1)}J_{-}^{(3)}-J_{-}^{(1)}J_{+}^{(3)})R^{(1)}R^{(2)}+R^{(1)}R^{(3)}/2-Q_{1}R^{(3)}-Q_{3}R^{(1)}.
\end{align*}
In the present realization the Casimir operator of the Bannai-Ito algebra becomes
\begin{align*}
Q_{\text{BI}}=\mu_1^2+\mu_2^2+\mu_3^2+\mu_4^2-1/4.
\end{align*}
It has been shown in section 3 that the Bannai-Ito polynomials form a basis for a representation of the BI algebra. It is here relatively easy to construct the representation of the BI algebra on bases of the three-fold tensor product module $V$ with basis vectors defined as eigenvectors of $Q_{12}$ or of $Q_{23}$. The first step is to obtain the spectra of the intermediate Casimir operators. Simple considerations based on the nature of the $sl_{-1}(2)$ representation show that the eigenvalues $q_{12}$ and $q_{23}$ of $Q_{12}$ and $Q_{23}$ take the form \cite{Genest-2013-12, Genest-2013-02, Genest-2012, Tsujimoto-2011-10}:
\begin{align*}
q_{12}=(-1)^{s_{12}+1}(s_{12}+\mu_1+\mu_2+1/2),\quad q_{23}=(-1)^{s_{23}}(s_{23}+\mu_2+\mu_3+1/2),
\end{align*}
where $s_{12}, s_{23}=0,1,\ldots, N$. The non-negative integer $N$ is specified by
\begin{align*}
N+1=\mu_4-\mu_1-\mu_2-\mu_3.
\end{align*}
Denote the eigenstates of $K_3$ by $\ket{k}$ and those of $K_1$ by $\ket{s}$; one has
\begin{align*}
K_3 \ket{k}=(-1)^{k}(k+\mu_1+\mu_2+1/2)\ket{k},\quad K_1\ket{s}=(-1)^{s}(s+\mu_2+\mu_3+1/2)\ket{s}.
\end{align*}
Given the expressions \eqref{Structure-Constants} for the structure constants $\Omega_{k}$, one can proceed to determine the $(N+1)\times (N+1)$ matrices that verify the anticommutation relations \eqref{BI-Racah}. The action of $K_1$ on $\ket{k}$ is found to be \cite{Genest-2012}:
\begin{align*}
K_1\ket{k}=U_{k+1}\ket{k+1}+V_{k}\ket{k}+U_{k}\ket{k-1},
\end{align*}
with $V_{k}=\mu_2+\mu_3+1/2-B_{k}-D_{k}$ and $U_{k}=\sqrt{B_{k-1}D_{k}}$ where
\begin{align*}
B_{k}&=
\begin{cases}
\frac{(k+2\mu_2+1)(k+\mu_1+\mu_2+\mu_3-\mu+1)}{2(k+\mu_1+\mu_2+1)} & \text{$k$ even,}
\\
\frac{(k+2\mu_1+2\mu_2+1)(k+\mu_1+\mu_2+\mu_3+\mu+1)}{2(k+\mu_1+\mu_2+1)} & \text{$k$ odd,}
\end{cases}
\\
D_{k}&=
\begin{cases}
-\frac{k(k+\mu_1+\mu_2-\mu_3-\mu)}{2(k+\mu_1+\mu)2)} & \text{$n$ even},
\\
-\frac{(k+2\mu_1)(k+\mu_1+\mu_2-\mu_3+\mu)}{2(k+\mu_1+\mu_2)} & \text{$n$ odd}.
\end{cases}
\end{align*}
Under the identifications
\begin{align*}
\rho_1=\frac{1}{2}(\mu_2+\mu_3),\quad \rho_2=\frac{1}{2}(\mu_1+\mu),\quad r_1=\frac{1}{2}(\mu_3-\mu_2),\quad r_2=\frac{1}{2}(\mu-\mu_1),
\end{align*}
one has $B_{k}=2A_{k}$, $D_{k}=2C_{k}$, where $A_{k}$ and $C_{k}$ are the recurrence coefficients \eqref{BI-RECU} of the Bannai-Ito polynomials. Upon setting 
\begin{align*}
\braket{s}{k}=w(s) 2^{k} B_k(x_{s}),\qquad B_{0}(x_{s})\equiv 1,
\end{align*}
one has on the one hand
\begin{align*}
\Braket{s}{K_1}{k}=(-1)^{s}(s+2\rho_1+1/2) \,\braket{s}{k},
\end{align*}
and on the other hand
\begin{align*}
\Braket{s}{K_1}{k}= U_{k+1} \braket{s}{k+1}+V_{k}\braket{s}{k}+U_{k-1}\braket{s}{k-1}.
\end{align*}
Comparing the two RHS yields
\begin{align*}
x_{s} B_{k}(x_{s})=B_{k+1}(x_{s})+(\rho_1-A_{k}-C_{k})B_{k}(x_{s})+A_{k-1}C_{k} B_{k-1}(x_{s}),
\end{align*}
where $x_{s}$ are the points of the Bannai-Ito grid
\begin{align*}
x_{s}=(-1)^{s}\left(\frac{s}{2}+\rho_1+1/4\right)-1/4,\quad s=0,\ldots, N.
\end{align*}
Hence the Racah coefficients of $sl_{-1}(2)$ are proportional to the Bannai-Ito polynomials. The algebra \eqref{BI-Racah} with structure constants \eqref{Structure-Constants} is invariant under the cyclic permutations of the pairs $(K_i, \mu_i)$. As a result, the representations in the basis where $K_1$ is diagonal can be obtained directly. In this basis, the operator $K_3$ is seen to be tridiagonal, which proves again that the Bannai-Ito polynomials possess the Leonard duality property.
\section{A superintegrable model on $S^2$ with Bannai-Ito symmetry}
We shall now use the analysis of the Racah problem for $sl_{-1}(2)$ and its realization in terms of Dunkl operators to obtain a superintegrable model on the two-sphere. Recall that a quantum system in $n$ dimensions with Hamiltonian $H$ is maximally superintegrable it it possesses $2n-1$ algebraically independent constants of motion, where one of these constants is $H$ \cite{Miller-2013-10}. Let $(s_1,s_2,s_3)\in \mathbb{R}$ and take $s_1^2+s_2^2+s_3^2=1$. The standard angular momentum operators are
\begin{align*}
L_1=\frac{1}{i}\left(s_2 \frac{\partial}{\partial s_3}-s_3 \frac{\partial}{\partial s_2}\right),\quad L_2=\frac{1}{i}\left(s_3 \frac{\partial}{\partial s_1}-s_1 \frac{\partial}{\partial s_3}\right),
\quad L_3=\frac{1}{i}\left(s_1 \frac{\partial}{\partial s_2}-s_2 \frac{\partial}{\partial s_1}\right).
\end{align*}
The system governed by the Hamiltonian
\begin{align}
\label{Hamiltonian}
H=L_1^2+L_2^2+L_3^2+\frac{\mu_1}{s_1^2}(\mu_1-R_1)+\frac{\mu_2}{s_2^2}(\mu_2-R_{2})+\frac{\mu_3}{s_3^2}(\mu_3-R_3),
\end{align}
with $\mu_i$, $i=1,2,3$, real parameters such that $\mu_i>-1/2$ is superintegrable \cite{Genest-2014-1}. 
\begin{enumerate}
\item The operators $R_i$ reflect the variable $s_i$: $R_i f(s_i)=f(-s_i)$.
\item The operators $R_i$ commute with the Hamiltonian: $[H, R_i]=0$.
\item If one is concerned with the presence of reflection operators in a Hamiltonian, one may replace $R_i$ by $\kappa_i=\pm 1$. This then treats the 8 potential terms
\begin{align*}
\frac{\mu_1}{s_1^2}(\mu_1-\kappa_1)+\frac{\mu_2}{s_2^2}(\mu_2-\kappa_2)+\frac{\mu_3}{s_3^2}(\mu_3-\kappa_3),
\end{align*}
simultaneously much like supersymmetric partners.
\item Rescaling $s_i\rightarrow r s_i$ and taking the limit as $r\rightarrow \infty$ gives the Hamiltonian of the Dunkl oscillator \cite{Genest-2013-04, Genest-2013-09}
\begin{align*}
\widetilde{H}=-[D_{x_1}^2+D_{x_2}^2]+\widehat{\mu}_3^2(x_1^2+x_2^2),
\end{align*}
after appropriate renormalization; see also \cite{Genest-2013-10, Genest-2013-07, Genest-2013-12-1}.
\end{enumerate}
It can be checked that the following three quantities commute with the Hamiltonian \eqref{Hamiltonian} \cite{Genest-2013-12, Genest-2014-1}:
\begin{align*}
C_1&=\left(i L_1+\mu_2 \frac{s_3}{s_2}R_2-\mu_3 \frac{s_2}{s_3} R_{3}\right)R_2+\mu_2 R_3+\mu_3 R_2+R_2R_3/2,
\\
C_2&=\left(-i L_2+\mu_1 \frac{s_3}{s_1} R_1-\mu_3 \frac{s_1}{s_3} R_{3}\right)R_1R_2+\mu_1 R_3+\mu_3 R_1+R_1R_3/2,
\\
C_3&=\left(i L_3+\mu_1 \frac{s_2}{s_1}R_1-\mu_2 \frac{s_1}{s_2}R_2\right)R_1+\mu_1 R_2+\mu_2 R_1+R_1R_2/2,
\end{align*}
that is, $[H, C_i]=0$ for $i=1,2,3$. To determine the symmetry algebra generated by the above constants of motion, let us return to the Racah problem for $sl_{-1}(2)$. Consider the following (gauge transformed) parabosonic realization of $sl_{-1}(2)$ in the three variables $s_i$:
\begin{align}
\label{Para-Realization}
J_{\pm}^{(i)}=\frac{1}{\sqrt{2}}\left[s_i \mp \partial_{s_i}\pm \frac{\mu_i}{s_i}R_i\right],\quad J_0^{(i)}=\frac{1,}{2}\left[-\partial_{s_i}^2+s_i^2+\frac{\mu_i}{s_i^2}(\mu_i-R_i)\right],\quad R^{(i)}=R_{i},
\end{align}
for $i=1,2,3$. Consider also the addition of these three realizations so that 
\begin{align}
\label{Realization-2}
\begin{aligned}
J_0=J_0^{(1)}+J_0^{(2)}+J_0^{(3)}, \quad J_{\pm}=J_{\pm}^{(1)}R^{(2)}R^{(3)}+J_{\pm}^{(2)}R^{(3)}+J_{\pm}^{(3)},\quad R=R^{(1)}R^{(2)}R^{(3)}.
\end{aligned}
\end{align}
It is observed that in the realization \eqref{Realization-2}, the total Casimir operator can be expressed in terms of the constants of motion as follows:
\begin{align*}
Q=-C_1 R^{(1)}-C_2 R^{(2)}-C_3 R^{(3)}+\mu_1 R^{(2)}R^{(3)}+\mu_2 R^{(1)}R^{(3)}+\mu_3 R^{(1)}R^{(2)}+R/2,
\end{align*}
Upon taking
\begin{align*}
\Omega=Q R,
\end{align*}
one finds
\begin{align}
\label{Expression}
\Omega^2+\Omega=L_1^2+L_2^2+L_3^2+(s_1^2+s_2^2+s_3^2)\left(\frac{\mu_1}{s_1^2}(\mu_1-R_1)+\frac{\mu_2}{s_2^2}(\mu_2-R_{2})+\frac{\mu_3}{s_3^2}(\mu_3-R_3)\right),
\end{align}
so that $H=\Omega^2+\Omega$ if $s_1^2+s_2^2+s_3^2=1$. Assuming this constraint can be imposed, $H$ is a quadratic combination of $QR$. By construction, the intermediate Casimir operators $Q_{ij}$ commute with the total Casimir operator $Q$ and with $R$ and hence with $\Omega$; they thus commute with $H=\Omega^2+\Omega$ and are the constants of motion. It is indeed found that
\begin{align*}
Q_{12}=-C_3,\qquad Q_{23}=-C_{1},
\end{align*}
in the parabosonic realization \eqref{Para-Realization}. Let us return to the constraint $s_1^2+s_2^2+s_3^2=1$. Observe that
\begin{align*}
\frac{1}{2}(J_{+}+J_{-})^2=(s_1 R_2 R_3+s_2 R_3+s_3)^2=s_1^2+s_2^2+s_3^2.
\end{align*}
Because $(J_{+}+J_{-})^2$ commutes with $\Omega=QR$, $Q_{12}$ and $Q_{23}$,  one can impose $s_1^2+s_2^2+s_3^2=1$. Since it is already known that the intermediate Casimir operators in the addition of three $sl_{-1}(2)$ representations satisfy the Bannai-Ito structure relations, the constants of motion verify
\begin{align*}
\{C_1,C_2\}&=C_3-2\mu_3 Q+2\mu_1 \mu_2,
\\
\{C_2, C_3\}&= C_1-2\mu_1 Q+2\mu_2 \mu_3,
\\
\{C_3, C_1\}&= C_2-2\mu_2 Q+2\mu_3 \mu_1,
\end{align*}
and thus the symmetry algebra of the superintegrable system with Hamiltonian \eqref{Hamiltonian} is a central extension (with $Q$ begin the central operator) of the Bannai-Ito algebra. Let us note that the relation $H=\Omega^2+\Omega$ relates to chiral supersymmetry since with $S=\Omega+1/2$ one has
\begin{align*}
\frac{1}{2}\{S,S\}=H+1/4.
\end{align*}
\section{A Dunkl-Dirac equation on $S^2$}
Consider the $\mathbb{Z}_2$-Dunkl operators
\begin{align*}
D_i=\frac{\partial}{\partial x_i}+\frac{\mu_i}{x_i}(1-R_i),\qquad i=1,2,\ldots,n,
\end{align*}
with $\mu_i>-1/2$. The $\mathbb{Z}_2^{n}$-Dunkl-Laplace operator is
\begin{align*}
\vec{D}^2=\sum_{i=1}^{n} D_i^2.
\end{align*}
With $\gamma_n$ the generators of the Euclidean Clifford algebra
\begin{align*}
\{\gamma_m,\gamma_{n}\}=2\delta_{nm},
\end{align*}
the Dunkl-Dirac operator is
\begin{align*}
\slashed{D}=\sum_{i=1}^{n} \gamma_i D_{i}.
\end{align*}
Clearly, one has $\slashed{D}^2=\vec{D}^2$. Let us consider the three-dimensional case. Introduce the Dunkl ``angular momentum'' operators
\begin{align*}
J_1=\frac{1}{i}(x_2 D_3-x_3 D_2),\quad  J_2=\frac{1}{i}(x_3 D_1-x_1 D_3),\quad  J_3=\frac{1}{i}(x_1 D_2-x_2 D_1).
\end{align*}
Their commutation relations are found to be
\begin{align}
\label{Comm-1}
[J_i, J_k]=i\epsilon_{jkl} J_{l}(1+2\mu_{l} R_{l}).
\end{align}
The Dunkl-Laplace equation separates in spherical coordinates; i.e. one can write
\begin{align*}
\vec{D}^2=D_1^2+D_2^2+D_3^2=\mathcal{M}_r+\frac{1}{r^2}\Delta_{S^2},
\end{align*}
where $\Delta_{S^2}$ is the Dunkl-Laplacian on the 2-sphere. It can be verified that \cite{Genest-2013-12-1}
\begin{align}
\label{J-Square}
\begin{aligned}
\vec{J}^2&=J_1^2+J_2^2+J_3^2
\\
&=-\Delta_{S^2}+2\mu_1\mu_2(1-R_1R_2)+2\mu_2 \mu_3 (1-R_2R_3)+2\mu_1 \mu_3 (1-R_1R_3)
\\
& \qquad \qquad \qquad  -\mu_1 R_1-\mu_2 R_2-\mu_3 R_3+\mu_1+\mu_2+\mu_3.
\end{aligned}
\end{align}
In three dimensions the Euclidean Clifford algebra is realized by the Pauli matrices
\begin{align*}
\sigma_1=
\begin{pmatrix}
0 & 1 
\\
1 & 0
\end{pmatrix},
\quad
\sigma_2=
\begin{pmatrix}
0 & -i
\\
i & 0
\end{pmatrix},
\quad
\sigma_3=
\begin{pmatrix}
1 & 0
\\
0 & -1
\end{pmatrix},
\end{align*}
which satisfy
\begin{align*}
\sigma_i\sigma_j=i \epsilon_{ijk}\sigma_k+\delta_{ij}.
\end{align*}
Consider the following operator:
\begin{align*}
\Gamma=(\vec{\sigma}\cdot \vec{J})+\vec{\mu}\cdot \vec{R},
\end{align*}
with $\vec{\mu}\cdot \vec{R}=\mu_1 R_1+\mu_2 R_2+\mu_3 R_3$. Using the commutation relations \eqref{Comm-1} and the expression \eqref{J-Square} for $\vec{J}^2$, it follows that
\begin{align*}
\Gamma^2+\Gamma=-\Delta_{S^2}+(\mu_1+\mu_2+\mu_3)(\mu_1+\mu_2+\mu_3+1).
\end{align*}
This is reminiscent of the expression \eqref{Expression} for the superintegrable system with Hamiltonian \eqref{Hamiltonian} in terms of the $sl_{-1}(2)$ Casimir operator. This justifies calling $\Gamma$ a Dunkl-Dirac operator on $S^2$ since a quadratic expression in $\Gamma$ gives $\Delta_{S^2}$. The symmetries of $\Gamma$ can be constructed. They are found to have the expression \cite{DeBie-2014}
\begin{align*}
M_i=J_i+\sigma_i(\mu_j R_j+\mu_k R_k+1/2),\quad \text{$(ijk)$ cyclic},
\end{align*}
and one has $[\Gamma, M_i]=0$. It is seen that the operators
\begin{align*}
X_i=\sigma_i R_i\qquad i=1,2,3
\end{align*}
also commute with $\Gamma$. Furthermore, one has
\begin{align*}
[M_i, X_i]=0,\qquad \{M_i, X_j\}=\{M_i, X_k\}=0.
\end{align*}
Note that $Y=-i X_1X_2X_3=R_1R_2R_3$ is central (like $\Gamma$). The commutation relations satisfied by the operators $M_i$ are
\begin{align*}
[M_i, M_j]=i\epsilon_{ijk}\left(M_{k}+2\mu_k(\Gamma+1)X_{k}\right)+2\mu_i\mu_j[X_i, X_j].
\end{align*}
This is again an extension of $\mathfrak{su}(2)$ with reflections and central elements. Let 
\begin{align*}
K_i=M_i X_i Y=M_i \sigma_i R_{j}R_{k}.
\end{align*}
It is readily verified that the operators $K_i$ satisfy
\begin{align*}
\{K_1,K_2\}&=K_3+2\mu_3(\Gamma+1)Y+2\mu_1\mu_2,
\\
\{K_2,K_3\}&=K_1+2\mu_1 (\Gamma+1)Y+2\mu_2\mu_3,
\\
\{K_3,K_1\}&= K_2+2\mu_3(\Gamma+1)Y+2\mu_3\mu_1,
\end{align*}
showing that the Bannai-Ito algebra is a symmetry subalgebra of the Dunkl-Dirac equation on $S^2$. Therefore, the Bannai-Ito algebra is also a symmetry subalgebra of the Dunkl-Laplace equation.
\section{Conclusion}
In this paper, we have presented the Bannai-Ito algebra together with some of its applications. In concluding this overview, we identify some open questions.
\begin{enumerate}
\item Representation theory of the Bannai-Ito algebra\medskip

Finite-dimensional representations of the Bannai-Ito algebra associated to certain models were presented. However, the complete characterization of all representations of the Bannai-Ito algebra is not known.
\item Supersymmetry\medskip

The parallel with supersymmetry has been underscored at various points. One may wonder if there is a deeper connection. 

\item Dimensional reduction \medskip

It is well known that quantum superintegrable models can be obtained by dimensional reduction. It would be of interest to adapt this framework in the presence of reflections operators. Could the BI algebra can be interpreted as a $W$-algebra ?\medskip

\item Higher ranks\medskip

Of great interest is the extension of the Bannai-Ito algebra to higher ranks, in particular for many-body applications. In this connection, it can be expected that the symmetry analysis of higher dimensional superintegrable models or Dunkl-Dirac equations will be revealing.
\end{enumerate}
\section*{Acknowledgements}
V.X.G. holds an Alexander-Graham-Bell fellowship from the Natural Science and Engineering Research Council of Canada (NSERC). The research of L.V. is supported in part by NSERC. H. DB. and A.Z. have benefited from the hospitality of the Centre de recherches math\'ematiques (CRM).

\section*{References}
\bibliographystyle{iopart-num}
\providecommand{\newblock}{}

\end{document}